\documentclass[aps,prb,preprint,superscriptaddress,floatfix]{revtex4}

\usepackage{graphicx,color}

\newcommand{\jwj}[1]{\textcolor{red}{#1}}

\graphicspath{{figs/}}
\begin{document}
\title{Negative Poisson's Ratio in Single-Layer Black Phosphorus}
\author{Jin-Wu Jiang  \footnote{Correspondence to: jiangjinwu@shu.edu.cn (JWJ)}}
    \affiliation{Shanghai Institute of Applied Mathematics and Mechanics, Shanghai Key Laboratory of Mechanics in Energy Engineering, Shanghai University, Shanghai 200072, People's Republic of China}
\author{Harold S. Park \footnote{Correspondence to: parkhs@bu.edu (HSP)}}
    \affiliation{Department of Mechanical Engineering, Boston University, Boston, Massachusetts 02215, USA}

\date{\today}
\begin{abstract}

\jwj{The Poisson's ratio is a fundamental mechanical property that relates the resulting lateral strain to applied axial strain.  While this value can theoretically be negative, it is positive for nearly all materials, though negative values have been observed in so-called auxetic structures.  However, nearly all auxetic materials are bulk materials whose microstructure has been specifically engineered to generate a negative Poisson's ratio.  Here, we report using first principles calculations the existence of a negative Poisson's ratio in a single-layer, two-dimensional material, black phosphorus.  In contrast to engineered bulk auxetics, this behavior is intrinsic for single layer black phosphorus, and originates from its puckered structure, where the pucker can be regarded as a re-entrant structure that is comprised of two coupled orthogonal hinges.  As a result of this atomic structure, a negative Poisson's ratio is observed in the out-of-plane direction under uniaxial deformation in the direction parallel to the pucker.}

\end{abstract}

\maketitle
\pagebreak

Nearly all materials have a positive Poisson's ratio, which means that when stretched longitudinally, they become shorter in the lateral directions.  In contrast, the list of materials that exhibit a negative Poisson's ratio is small, and thus researchers have shown significant interest in finding new materials that exhibit this distinctive property.  Negative Poisson's ratio materials are also of great interest because they typically have enhanced toughness and shear resistance, along with enhanced sound and vibration absorption, which has been exploited in fields as varied as medicine\cite{ScarpaF2008}, fasteners,\cite{ChoiJB1991cp} tougher composites,\cite{SunY2013m} tissue engineering,\cite{ParkYJ2013amse} national security and defense, i.e. for bulletproof vests and armor enhancement\cite{LiuQ2006}, and many other potential applications.\cite{PrawotoY2012cms,EvansKE2000am,YangW2004jms}

Most reports of negative Poisson's ratio have been from bulk-scale, engineered structures.  For example, the first foam with a negative Poisson's ratio was reported in 1987 by Lakes\cite{LakesRS1987sci}; such materials that exhibit a negative Poisson's ratio were eventually termed auxetics by Evans et al. in 1991.\cite{EVANSKE1991}  The negative Poisson's ratio in these bulk structures is derived from engineering of the material microstructure.\cite{LakesR1991jmsci}  For example, the negative Poisson's ratio in the foam of Lakes\cite{LakesRS1987sci} is due to a particular microstructure containing a re-entrant unit cell with two orthogonal hinges.\cite{FriisEA1988jms} During the mechanical tension of a unit cell in the foam, one of the hinges  opens along the tensile axis, while the other hinge expands in the lateral direction.  This re-entrant mechanism  can result in a negative Poisson's ratio in auxetic foams, though we note that negative Poisson's ratio materials do not require hinges.

While bulk materials with engineered microstructure have been the focus for generating materials exhibiting a negative Poisson's ratio, this is not a requirement for the phenomenon to occur.  For example, negative Poisson's ratio has been observed in cubic materials \cite{MilsteinF,BaughmanRH1998nat} and also crystalline SiO$_{2}$.\cite{KeskarNR1992nat} Many theoretical studies of auxetics have been performed on a variety of model systems, for example as done by Wojciechowski\cite{WojciechowskiKW1989pla}, while a Poisson's ratio of -1 was found in chiral honeycomb systems.\cite{PrallD1996ijms}  Furthermore, a negative Poisson's ratio was also found in auxetic origami sheets.\cite{GreavesGN2011nm,WeiZY2013prl}  Ho et al. recently reported that negative Poisson's ratios can be found in metal nanoplates due to a combination of free surface effects, and a phase transformation under uniaxial loading.\cite{HoDT2014nc}

While significant effort has been expended in studying negative Poisson's ratio in various materials and structures, we are not aware of such a report in one-atom-thick, or two-dimensional materials such as graphene or boron nitride, which have already been the focus of extensive research over the past decade due to their unique mechanical, electronic, thermal, optical and other physical properties.\cite{GeimAK2007nm,LiD2008sci,GeimAK2009sci,CastroNAH,RaoCNR,Balandin2011nm,NovoselovKS2005pnas} Therefore, we are interested in examining possible negative Poisson's ratio effects in these two-dimensional nanomaterials.

In this work, we focus for three reasons on a recent arrival to the world of two-dimensional materials, single-layer black phosphorus (BP).  First, few-layer BP has recently been explored as an alternative electronic material to graphene, boron nitride, and the transition metal dichalcogenides for transistor applications\cite{LiL2014,LiuH2014,BuscemaM2014}.  This is because single-layer BP has a direct bandgap that can be modulated through thickness and strain,\cite{LiuH2014,DuY2010jap,RodinAS2014,Castellanos-GomezA2014arxiv} and exhibits a high carrier mobility\cite{LiuH2014} and large optical conductivity.\cite{XiaF2014} Furthermore, single-layer BP has a characteristic puckered structure as shown in Fig.~\ref{fig_cfg}~(a), which should lead to substantial anisotropy in the mechanical behavior and properties, i.e. the Poisson's ratio, along both the two in-plane and the out-of-plane directions.  Finally, the pucker can be thought of as consisting of two orthogonal hinges $\theta_{546}$ and $\theta_{214}$ (or $\theta_{314}$).  Because these two coupling hinges strongly resemble those seen in the re-entrant unit cell that enable the negative Poisson's ratio in auxetic foams,\cite{LakesRS1987sci,FriisEA1988jms} we wish to investigate their impact on the Poisson's ratios in single-layer BP.

We report using first principles calculations the existence of a negative Poisson's ratio in a single-layer, two-dimensional material, black phosphorus.  In contrast to previous reports on negative Poisson's ratio in bulk auxetic materials with specifically engineered microstructures\cite{MilsteinF,ChoiJB1992jms} or more recently in metal nanoplates due to surface and large deformation effects,\cite{HoDT2014nc} this behavior is intrinsic for single layer black phosphorus, and originates from its puckered structure, where the pucker can be regarded as a re-entrant structure that is comprised of two coupled orthogonal hinges.  As a result of this atomic structure, a negative Poisson's ratio is observed in the out-of-plane direction under uniaxial deformation in the direction parallel to the pucker, which becomes increasingly negative with both increased tension and compression. 

\section*{Results}
\begin{figure}[htpb]
  \begin{center}
    \scalebox{1.0}[1.0]{\includegraphics[width=8cm]{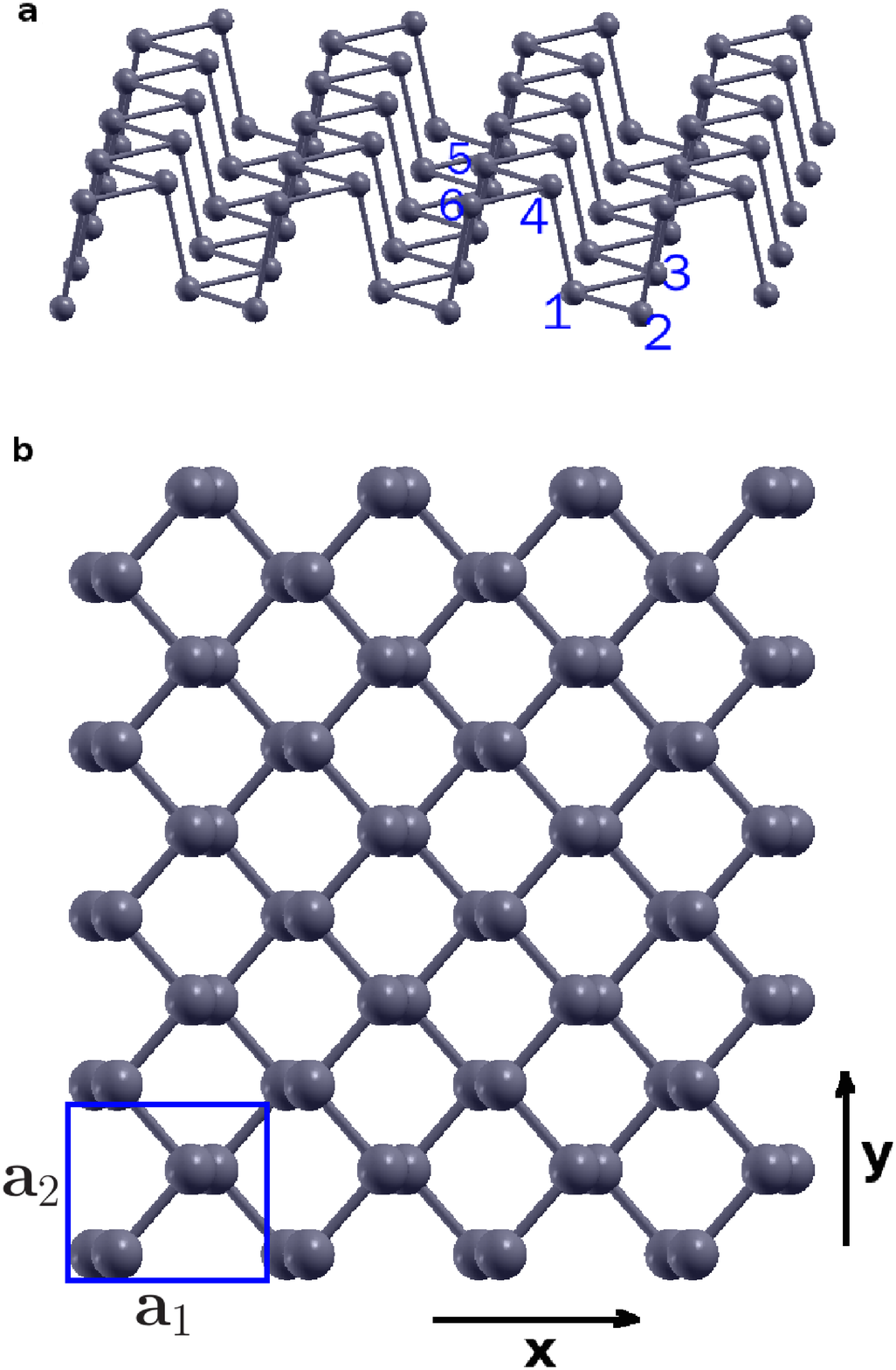}}
  \end{center}
  \caption{(Color online) Optimized configuration of single-layer BP. (a) Perspective view illustrates the pucker along the y-direction. (b) Top view of top image showing a honeycomb-like structure.  The blue box represents the basic unit cell for single-layer BP.}
  \label{fig_cfg}
\end{figure}

\textbf{Structural stability.} The structure of the single-layer BP is optimized using the first-principles calculations; the computational details are given in the Methods section. Fig.~\ref{fig_cfg} shows via plotting with XCRYSDEN\cite{xcrysden} the relaxed structure for a single-layer of BP of dimension $17.69\times16.74$~{\AA}. Fig.~\ref{fig_cfg}~(a) shows a perspective view that displays the puckered configuration of single-layer BP. In this puckered structure, each P atom is connected to three neighboring P atoms. There are two inequivalent P-P bonds in the relaxed structure, i.e $r_{12}=r_{13}=2.4244$~{\AA} and $r_{14}=2.3827$~{\AA}, and also two inequivalent bond angles $\theta_{213}=98.213^{\circ}$ and $\theta_{214}=\theta_{314}=97.640^{\circ}$. The blue box in Fig.~\ref{fig_cfg}~(b) shows the unit cell with four P atoms. The two lattice constants are $a_{1}=4.1319$~{\AA} and $a_{2}=3.6616$~{\AA}. These structural parameters are close to the experimental values.\cite{BrownA1965ac}  The top view depicted in Fig.~\ref{fig_cfg}~(b) shows a honeycomb-like lattice structure for single-layer BP. Note that two edges in the honeycomb are much shorter than the other four edges.  The Cartesian coordinates are set with the x-direction perpendicular to the pucker and the y-direction parallel with the pucker.

The geometric anisotropy introduced by the pucker implies that single-layer BP may exhibit significant anisotropy in the mechanical response to uniaxial loading in the x and y-directions. Because the mechanical behavior of single-layer BP has not been reported in the literature, and to illustrate the anisotropic deformation that is related to the existence of the negative Poisson's ratio, we first discuss the changes in atomic structure that results in single-layer BP from uniaxial deformation in the both the x and y-directions.

\begin{figure}[htpb]
  \begin{center}
    \scalebox{1.7}[1.7]{\includegraphics[width=8cm]{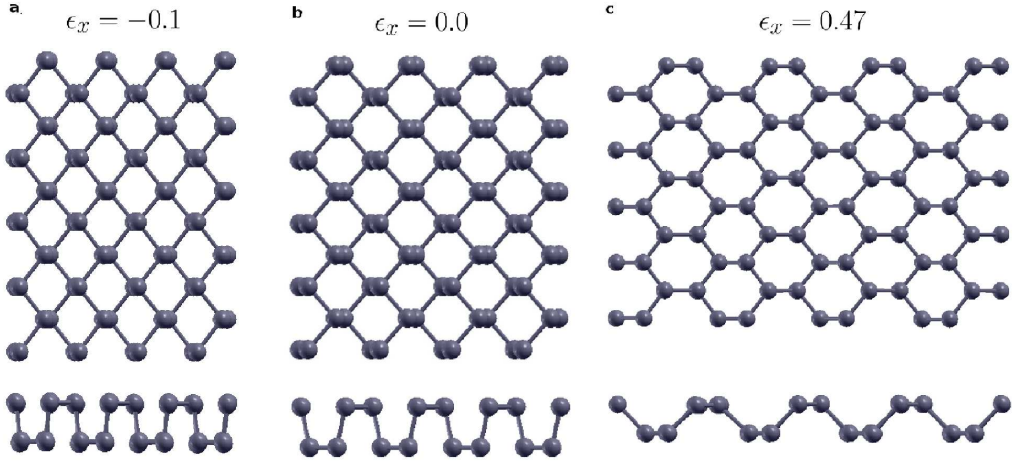}}
  \end{center}
  \caption{(Color online) The configuration of single-layer BP under uniaxial deformation in the x-direction. (a) Compression with $\epsilon_{x}=-0.1$. (b) Optimized structure without deformation ($\epsilon_{x}=0.0$). (c) Tension with $\epsilon_{x}=0.47$. Top and side views are displayed in all panels. Note the unfolding of the pucker in the side view in panel (c).}
  \label{fig_x_cfg}
\end{figure}

\begin{figure}[htpb]
  \begin{center}
    \scalebox{1.7}[1.7]{\includegraphics[width=8cm]{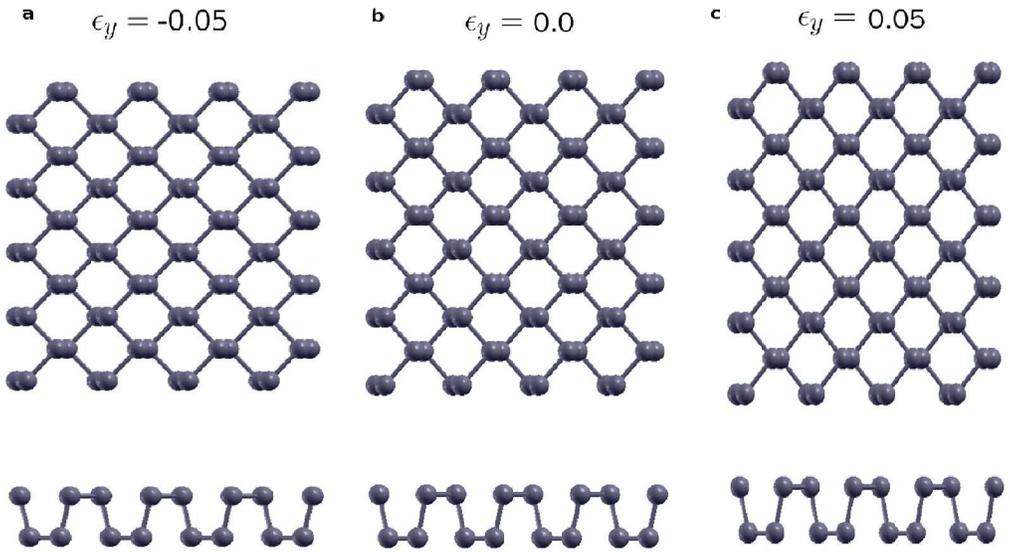}}
  \end{center}
  \caption{(Color online) The configuration of single-layer BP under uniaxial deformation in the y-direction. (a) Compression with $\epsilon_{y}=-0.05$. (b) Optimized structure without deformation ($\epsilon_{y}=0.0$). (c) Tension with $\epsilon_{y}=0.05$. Top and side views are displayed in all panels.}
  \label{fig_y_cfg}
\end{figure}

We compare the initial relaxed configuration for single-layer BP with its deformed configuration for loading both parallel and perpendicular to the pucker.  Fig.~\ref{fig_x_cfg} shows results for uniaxial tension and compression in the x-direction (perpendicular to the pucker).  Specifically, Fig.~\ref{fig_x_cfg} shows the compressed structure with $\epsilon_{x}=-0.1$ in panel (a), the initial relaxed (strain-free) configuration in panel (b), and the stretched configuration with $\epsilon_{x}=0.47$ in panel (c). The side views clearly show the significant unfolding of the puckers during tension.  Corresponding to this unfolding, the two shorter bonds in the top view become longer, and a regular honeycomb structure is formed as shown in the top view. Similarly, Fig.~\ref{fig_y_cfg} shows results for uniaxial tension and compression in the y-direction, where the compressed single-layer BP configuration with $\epsilon_{y}=-0.05$ is shown in panel (a), the initial relaxed (strain-free) structure is shown in panel (b), and the stretched configuration with $\epsilon_{y}=0.05$ is shown in panel (c).

\begin{figure}[htpb]
  \begin{center}
    \scalebox{1.7}[1.7]{\includegraphics[width=8cm]{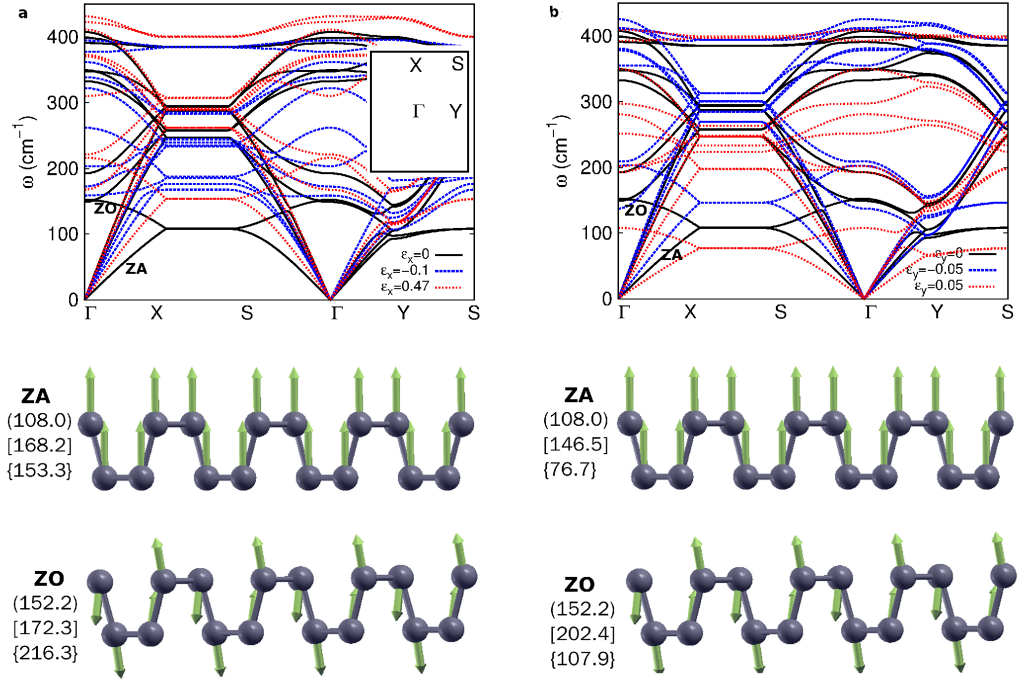}}
  \end{center}
  \caption{(Color online) Phonon dispersion for strained black phosphorus. (a) Top: Phonon dispersion for black phosphorus with strain $\epsilon_{x}=$ 0.0 (solid black lines), -0.1 (dotted blue lines), and 0.47 (dashed red lines). Inset shows the first Brillouin zone. Bottom: Eigenvector of z-direction acoustic (ZA) and z-direction optical (ZO) modes. Arrows attached on each atom illustrate the vibrational component of the atom in this phonon mode.  The three numbers on the left side are the frequency of the ZA mode (at X point) and ZO mode (at $\Gamma$ point) for black phosphorus with strain $\epsilon_{x}=$ 0.0, -0.1, and 0.47. (b) Top: Phonon dispersion for black phosphorus with strain $\epsilon_{y}=$ 0.0 (solid black lines), -0.05 (dotted blue lines), and 0.05 (dashed red lines). Bottom: Eigenvector of z-direction acoustic (ZA) and z-direction optical (ZO) modes.   The three numbers on the left side are the frequency of the ZA mode (at X point) and ZO mode (at $\Gamma$ point) for black phosphorus with strain $\epsilon_{y}=$ 0.0, -0.05, and 0.05.}
  \label{fig_zmode}
\end{figure}

To examine the stability of the strained structure shown in Figs.~\ref{fig_x_cfg} and \ref{fig_y_cfg}, we perform phonon dispersion calculations, which are widely used to examine the stability of deformed structures, of the strained single-layer BP.  If there is an imaginary mode with $\omega^2=-\gamma^2$ in the phonon dispersion, we get $\omega= \pm i\gamma$. For the imaginary mode with $\omega=i\gamma$ (with $\gamma>0$), the vibrational displacement ($u$) of some atoms will be divergent after a long time ($t\longrightarrow +\infty$),
\begin{eqnarray}
u\propto e^{-i\omega t} = e^{\gamma t} \longrightarrow +\infty.
\label{eq_u}
\end{eqnarray}
Fig.~\ref{fig_zmode}~(a) shows the phonon dispersion for black phosphorus when strain is applied in the $x$-direction, for strain values of $\epsilon_{x}=$ 0.0 (solid black lines), -0.1 (dotted blue lines), and 0.47 (dashed red lines). Fig.~\ref{fig_zmode}~(a) clearly shows that there is no imaginary mode in the strained black phosphorus; i.e these configurations are stable. Similarly, Fig.~\ref{fig_zmode}~(b) shows the phonon dispersion for single-layer BP with strain $\epsilon_{y}=$ 0.0 (solid black lines), -0.05 (doted blue lines), and 0.05 (dashed red lines). There is no imaginary phonon mode in these phonon dispersion either, so these configurations are also stable.

The phonon calculation also reveals some interesting phenomena for the z-direction acoustic (ZA) and z-direction optical (ZO) modes. The eigenvectors of these modes are shown in both panels (a) and (b). Arrows attached on each atom illustrate the vibrational component of the atom in this phonon mode. We find that x-directional strain and y-directional strain have different effects on the frequency of ZA and ZO modes. For x-directional strain, both compression and tension leads to the enhancement of the frequency for these two modes. However, for y-directional strain, the compression increases the frequency of ZA and ZO modes while tension reduces their frequencies.
To explain this difference, we calculate the strain energy density of the strained black phosphorus.  For x-directional strain with $\epsilon_{x}\in[-0.1, 0.47]$, the strain energy density can be fitted to the following function,
\begin{eqnarray}
E = \frac{29}{2}\epsilon_{x}^2 - \frac{110}{6}\epsilon_{x}^{3},
\label{eq_Ex}
\end{eqnarray}
where the cubic term accounts for strong nonlinear effects for large $\epsilon_{x}$. This cubic nonlinear term leads to a quadratic force, resulting in the same effect from both compression and tension, and explains why both compression and tension enhance the frequency of ZA and ZO modes.  However, for y-directional strain with $\epsilon_{y}\in[-0.05, 0.05]$, the strain energy density can be fitted to the following function,
\begin{eqnarray}
E = \frac{125}{2}\epsilon_{y}^2.
\label{eq_Ey}
\end{eqnarray}
This quadratic strain energy results in a linear force-displacement or stress-strain relationship, which depends on the direction of the mechanical strain, i.e compression and tension leads to opposite effect. This is the origin for the opposite effects from the compression and tension on the frequency of the ZA and ZO modes.

\begin{figure}[htpb]
  \begin{center}
    \scalebox{1.4}[1.4]{\includegraphics[width=8cm]{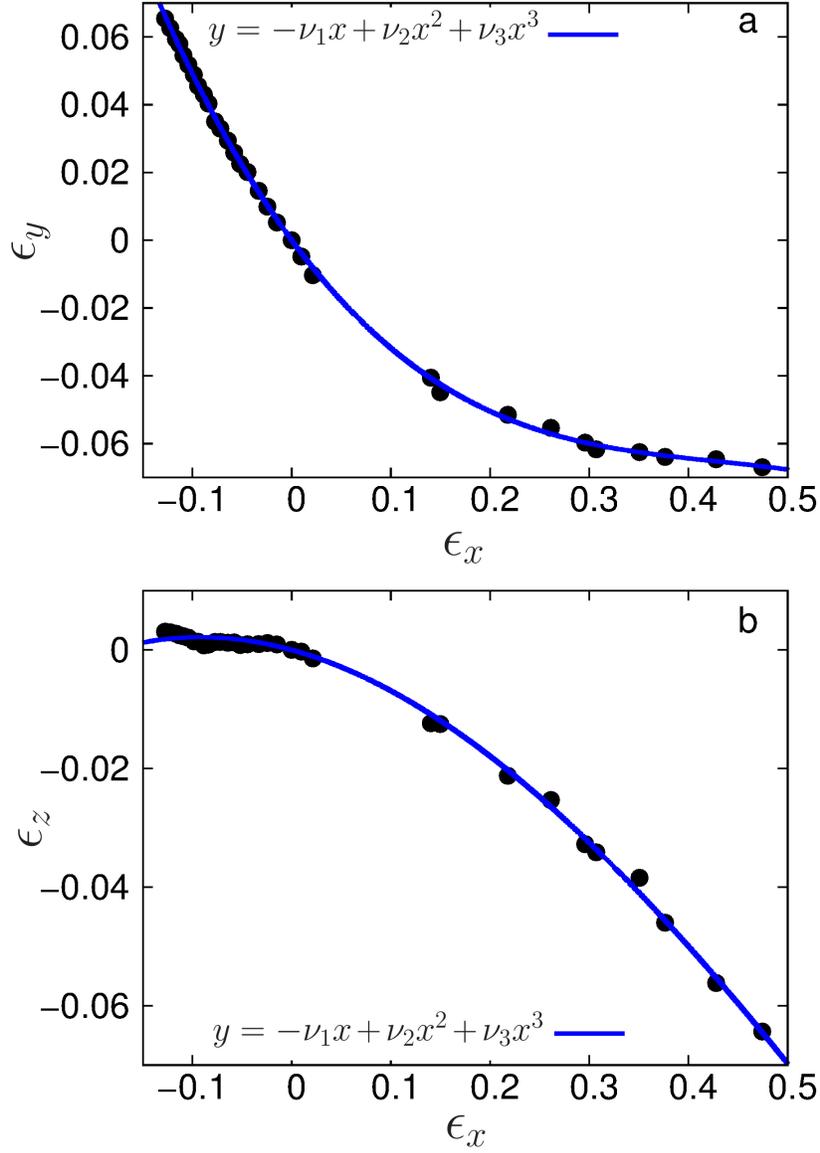}}
  \end{center}
  \caption{(Color online) The Poisson's ratio as a function of uniaxial deformation of single-layer BP in the x-direction. (a) $\epsilon_{y}$ versus $\epsilon_{x}$. Data are fitted to function $y=-\nu_{1} x + \nu_{2}x^2 + \nu_{3}x^3$, with $\nu_{1}=0.40$ as the linear Poisson's ratio, $v_{2}=0.86$ and $v_{3}=-0.67$. (b) $\epsilon_{z}$ versus $\epsilon_{x}$. Data are fitted to function $y=-\nu_{1} x + \nu_{2}x^2 + \nu_{3}x^3$, with $\nu_{1}=0.046$ as the linear Poisson's ratio, $v_{2}=-0.24$ and $v_{3}=0.10$.}
  \label{fig_poisson_x}
\end{figure}

\textbf{Poisson's ratio.} Having qualitatively illustrated how single-layer BP responds structurally to uniaxial deformation in the x and y-directions, we now focus on the resulting Poisson's ratios. The Poisson's ratios of single-layer BP are shown in Fig.~\ref{fig_poisson_x} for uniaxial deformation in the x-direction (perpendicular to the pucker). Fig.~\ref{fig_poisson_x}~(a) shows the strain in the y-direction during its deformation in the x-direction. A strongly nonlinear behavior is observed in the figure, especially in its tensile deformation region.  The data are fitted to the function $y=-\nu_{1} x + \nu_{2}x^2 + \nu_{3}x^3$, where $\nu=\nu_{1}=0.40$ can be regarded as the linear Poisson's ratio, which we define to be the value for infinitesimal deformations, or excluding nonlinear contributions. Similarly, Fig.~\ref{fig_poisson_x}~(b) shows that the linear Poisson's ratio in the z (out-of-plane) direction is  $\nu=\nu_{1}=0.046$.

\begin{figure}[htpb]
  \begin{center}
    \scalebox{1.5}[1.5]{\includegraphics[width=8cm]{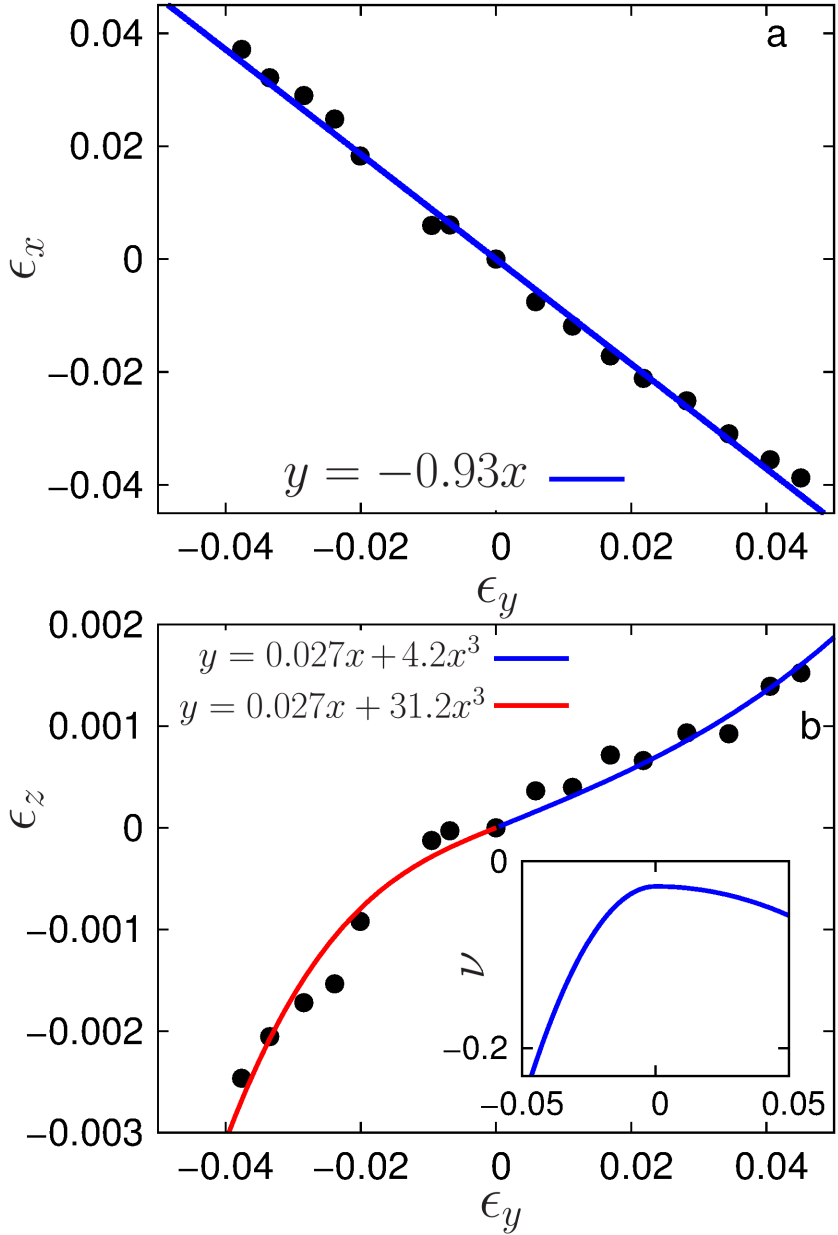}}
  \end{center}
  \caption{(Color online) The Poisson's ratio as a function of uniaxial deformation of single-layer BP in the y-direction. (a) $\epsilon_{x}$ versus $\epsilon_{y}$. Data are fitted to function $y=-\nu x$, with $\nu=0.93$ as the linear Poisson's ratio. (b) $\epsilon_{z}$ versus $\epsilon_{y}$. Data are fitted to function $y=-\nu_{1} x + \nu_{3}x^3$, with $\nu=\nu_{1}=-0.027$ as the linear Poisson's ratio in the out-of-plane direction. Inset shows the out-of-plane Poisson's ratio obtained from the fitting curves in panel (b) through $\nu = -\partial \epsilon_{z} / \partial\epsilon_{y}$. We note the negative Poisson's ratio for the full strain range.}
  \label{fig_poisson_y}
\end{figure}

Fig.~\ref{fig_poisson_y} provides information for the Poisson's ratio of single-layer BP due to uniaxial deformation in the y-direction (parallel to the pucker). Fig.~\ref{fig_poisson_y}~(a) shows that the linear Poisson's ratio in the y-direction is $\nu=0.93$, which is about two times larger than the Poisson's ratio in the x-direction. Hence, the linear Poisson's ratio is highly anisotropic in single-layer BP. Up to now, we have seen that single-layer BP will contract (expand) in the other two lateral directions, when it is stretched (compressed) in the x or y-direction. However, Fig.~\ref{fig_poisson_y}~(b) shows that single-layer BP expands (contracts) in the z-direction when it is stretched (compressed) in the y-direction; we can calculate that the negative linear Poisson's ratio is $\nu=-0.027$ in the out-of-plane direction during its deformation in the y-direction.  The figure inset further demonstrates that the out-of-plane Poisson's ratio is negative for the entire strain range we have considered, with the Poisson's ratio becoming increasingly negative with both increased tension and compression. We note that such negative out-of-plane Poisson's ratios have been found in some animal skins.\cite{LeesC1991bme} \jwj{We note that our applied y-directional strain range of -5\% to 5\% is actually above that achieved experimentally for various two-dimensional materials, which typically cannot exceed 3-4\%.\cite{HuangM2010nl}  The negative Poisson's ratio we report should thus be observable experimentally because it is negative for any y-directional strain value.}

\begin{figure}[htpb]
  \begin{center}
    \scalebox{1.6}[1.6]{\includegraphics[width=8cm]{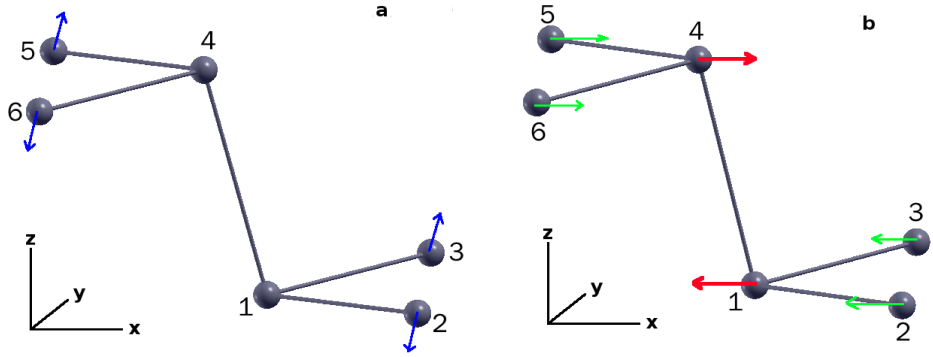}}
  \end{center}
  \caption{(Color online) The evolution of local structure in single-layer BP during uniaxial tension in the y-direction. (a) BP is stretched in the y-direction, i.e atoms are moved in the direction of the attached arrows (blue online). (b) To accommodate the tension in the y-direction, BP contracts in the x-direction, i.e atoms 1 and 4 move inward along the attached arrows (red online). The 1-4 bond thus becomes more closely aligned with the vertical (z)-direction.  The green arrows display the movement of the four surrounding atoms following the movement of atoms 1 and 4.}
  \label{fig_cfg_poisson}
\end{figure}

\textbf{Mechanistic explanation for the negative Poisson's ratio.} We offer two explanations to explain the origin of the negative Poisson's ratio, starting with a mechanistic summary.  Fig.~\ref{fig_cfg_poisson} illustrates the relationship between the puckered structure and the negative Poisson's ratio, where the motion of atoms for a representative strain increment in the y-direction is illustrated. Specifically, single-layer BP is stretched in the y-direction, i.e the four surrounding atoms are moved along the attached arrows (blue online) in Fig.~\ref{fig_cfg_poisson}~(a). The in-plane angles $\theta_{546}$ and $\theta_{213}$ increase and strain energy is stored in these two angles.  To accommodate the elongation in the y-direction, single-layer BP contracts in the x-direction, i.e atoms 1 and 4 move inward along the x-direction along the attached arrows (red online) in Fig.~\ref{fig_cfg_poisson}~(b), while the four surrounding atoms 2,3,5 and 6 follow the movement of atoms 1 and 4 (see green arrows). Furthermore, we note that due to the large Poisson's ratio ($\nu=0.93$) in the y-direction, single-layer BP will undergo substantial contraction in the x-direction during tension in the y-direction.  The inward movement of atoms 1 and 4 reduces the two in-plane angles $\theta_{546}$ and $\theta_{213}$, which were increased by the applied mechanical tension in the y-direction. As a consequence, the 1-4 bond becomes more closely aligned with the vertical (z)-direction, which alters the out-of-plane bond angles $\theta_{214}$ and $\theta_{314}$, and which causes strain energy to be shifted from the in-plane angles to the out-of-plane angles via the arrangement of bond 1-4. For $\epsilon_{y}=0.05$, we find $\theta_{214}=\theta_{314}=96.02^{\circ}$, which is smaller than the initial value ($97.640^{\circ}$). As a consequence of the movement of atoms 1 and 4 (red online), the single-layer BP thickness (i.e projection of bond length $r_{14}$ to the z-direction) is increased, and single-layer BP expands in the z-direction, leading to a negative Poisson's ratio in this direction.

\textbf{Re-entrant mechanism for the negative Poisson's ratio.} We have established that the negative Poisson's ratio in single-layer BP emerges due to its puckered atomic structure.  Structurally, we note that the pucker can be regarded as two coupling hinges formed by the angles $\theta_{546}$ and $\theta_{214}$ (or $\theta_{314}$). These two hinges are orthogonal to each other. When the hinge $\theta_{546}$ is opened in the y-direction, the other hinge $\theta_{214}$ (or $\theta_{314}$) closes along the orthogonal (x)-direction, which results in the expansion and negative Poisson's ratio in the out-of-plane (z)-direction. It is clear that the expansion in the z-direction directly results from the fact that $\theta_{214}>90^{\circ}$ and $\theta_{314}>90^{\circ}$. This coupling hinge mechanism is similar to the re-entrant mechanism previously exploited in auxetic foam.\cite{LakesRS1987sci,FriisEA1988jms} In the re-entrant foam, the idealized unit cell is also built based on coupling orthogonal hinges. When one of the hinges is opened in a particular direction, the other orthogonal hinge will expand in a lateral direction, leading to negative Poisson's ratio. In this sense, the pucker in single-layer BP functions as a nanoscale re-entrant structure. We thus expect that negative Poisson's ratios will be intrinsic to other puckered two-dimensional nanomaterials.

\section*{Discussion}
In conclusion, we have performed first-principles calculations on a single-layer of black phosphorus to reveal for the first time a negative Poisson's ratio in a two-dimensional nanomaterial.  The negative Poisson's ratio, which occurs in the out-of-plane direction, occurs due to the puckered atomic structure of single-layer black phosphorus.  This phenomena should be intrinsic to other puckered two-dimensional nanomaterials due to the fact that the pucker functions as a nanoscale re-entrant structure, which has previously been exploited in bulk materials to induce a negative Poisson's ratio.  We close by noting that two-dimensional auxetic materials may lead to novel material functionality as compared to those seen in bulk auxetics due to the unique material properties and functionalities that are intrinsic to two-dimensional nanomaterials.

\section*{Methods}
\textbf{First-principles calculation technique.} In this section, we detail the computational methodology used in this work. For the {\it ab initio} calculations, we used the SIESTA package\cite{SolerJM} to optimize the structure of single-layer BP. The generalized gradient approximation (GGA) was applied to account for the exchange-correlation function with Perdew-Burke-Ernzerhof (PBE) parametrization\cite{PerdewJP1996prl} and the double-$\zeta$ basis set orbital was adopted. It has been shown that the GGA is more accurate than the local density approximation for describing the structural and mechanical properties of single-layer BP.\cite{AppalakondaiahS2012prb} During the conjugate-gradient optimization, the maximum force on each atom is smaller than 0.01 eV\AA$^{-1}$. A mesh cut off of 120 Ry was used. Periodic boundary conditions were applied in the two in-plane transverse directions, while free boundary conditions were applied to the out-of-plane direction by introducing sufficient vacuum space of 15~{\AA}. Gamma point $k$ sampling was adopted for wave vector space.

\textbf{Force constant calculation technique.} We used the finite difference method to extract the force constant matrix from the first-principles calculations.\cite{YinMT1982prb} More specifically, each atom i is displaced in the $\alpha$ (= $\pm x$, $\pm y$, $\pm z$) direction for a small value of $\Delta_{i\alpha}=\Delta=0.04$~{Bohr}, which gives the force constant $C_{i\alpha;j\beta}$,
\begin{eqnarray}
C_{i\alpha;j\beta}=-\frac{F_{i\alpha}}{\Delta_{j\beta}},
\label{eq_c}
\end{eqnarray}
where $F_{i\alpha}$ is the force of atom i in the $\alpha$ direction due to the displacement of atom j in the $\beta$ direction. The force constants in the $x$, $y$, or $z$ direction are obtained from the average of the force constants corresponding to $\pm x$, $\pm y$, or $\pm z$. The dynamical matrix is constructed based on these force constant matrices. The diagonalization of the dynamical matrix results in the phonon frequency and eigenvector, i.e phonon dispersion.

\textbf{Acknowledgements} The work is supported by the Recruitment Program of Global Youth Experts of China and the start-up funding from the Shanghai University. HSP acknowledges the support of the Mechanical Engineering department at Boston University.

\textbf{Author contributions} J.W.J performed the calculations and discussed the results with H.S.P. J.W.J and H.S.P co-wrote the paper.

\textbf{Competing financial interests} The authors declare no competing financial interests.


\end{document}